\documentclass[a4paper,dvips 14pt]{revtex4}
\usepackage{epsfig,graphics,graphicx,amsmath,amssymb}
\usepackage{xcolor}

\begin{document}
	\baselineskip=.6cm
\title{Neutrino oscillation with minimal length uncertainty relation via wave packet approach}

  \author{M.M. Ettefaghi}\email{mettefaghi@qom.ac.ir}

\affiliation{Department of Physics, University of Qom, Ghadir Blvd., Qom 371614-611, I.R. Iran}

\begin{abstract} 	
Theories of Quantum Gravity as well as string theory suggest the existence  of a minimal measurable length  and the related Generalized Uncertainty
Principle (GUP). The universality of Quantum Gravity implies that the GUP influences every quantum mechanical process. Neutrino oscillation as a quantum phenomenon exhibits quantumness at macroscopic distances and could provide potentially a suitable room for quantum foundation explorations. In this paper, we perturbatively derive the neutrino oscillation probability based on the GUP and by treating neutrinos as wave packets. We see that the GUP modifications are dependent on the effective position width of the transition amplitude $\sigma_x$ such that with the smaller $\sigma_x$ we can obtain a stronger bound on the minimal length scale in comparison to what is expected from standard model interactions. More explicitly, one can obtain an upper bound about $5\times 10^{25}$ for the deformation parameter, $\beta_0$, with accelerator neutrino experiments such as MINOS, provided that $\sigma_x\sim 10^{-15}\text{m}$ which is reasonable since the energy of these neutrinos is of the order of a few GeV.  
\end{abstract}
\maketitle

\section{Introduction}\label{1}
At least at energy scales as large as the Planck scale, gravity as a fundamental interaction must be compatible with quantum mechanics. However, the quantization of gravity using the quantum field prescriptions leads to a non-renormalizable theory. Hence,  it has been believed that in the theories of quantum gravity there may exist an effective cutoff in the ultraviolet or equivalently a minimal measurable distance at the Planck scale \cite{gravity1,gravity2,gravity3}. Indeed, the belief that the minimal length exists is the result of a conjecture that gravity causes the structure of space-time to be disturbed at the Plank scale. Although, at the electroweak scale the effects of gravity interactions are so weak that they cannot be measured directly by recent and upcoming experiments, they are universal and anything which has energy creates gravity and is effected by it. The existence of a minimal measurable length, predicted by theories of Quantum Gravity, influences all quantum Hamiltonians \cite{das}. Therefore, it is suitable to consider these quantum gravity corrections to various quantum phenomena and obtain experimental bound on the scale at which these corrections appear.

 When there exists a minimal measurable length, the increase in energy cannot reduce the amount of uncertainty of the position measurement to arbitrary size. So the Heisenberg uncertainty principle, $\Delta x \Delta p \geq 1/2$,  breaks down for energies close the minimal length scale and it must be generalized  \cite{GUP1,GUP2,GUP3}. In one dimension, the Generalized Uncertainty Principle (GUP) is given by
 \begin{equation}
 	\Delta x\Delta p\geq \frac{1}{2}(1+\beta(\Delta p)^2+\beta\langle p\rangle^2),\label{gup}
\end{equation}
 where $\beta$ is a positive parameter defined as $\beta=\beta_0/M^2_{\text{Pl}}$ with $M_{\text{Pl}}\sim 10^{19} GeV$. If the dimensionless parameter $\beta_0$ is of the order of unity,  the new physics coming from minimal length correction cannot be measured even in future experiments. However, the hypothesis of extra dimensions paved the way for the investigation of quantum gravity at the achievable energy at the LHC \cite{ed1,ed2,ed3,ed4}. For instance, 	
 	according to Arkani-Hamed, Dimopolous and Dvali (ADD) model \cite{ed1}, the gauge interactions are confined to a 3-brane while gravity propagates
 	in all the dimensions. In this case, the effective scale of the appearance of quantum gravitational interactions $M_S$ is related to the usual Planck scale by $M_{Pl}^2=R^dM_S^{d+2}$ where $R$ is the radius of compactification and $d$ is the number of extra dimensions. Thus it is possible to choose $R$ for a given $d$ such that the quantum gravity scale comes down to	$M_S\sim 1$ TeV. Hence, it was proposed that high energy particle colliders could turn out to be huge BH factories \cite{bh1,bh2,bh3}. Therefore, the production of a microscopic black hole has been explored by the LHC data; the CMS collaboration excluded the production of a microscopic black hole with masses below 4.3
 	to 6.2 TeV (depending on the models) with 95 \% confidence
 	level \cite{cms} and the ATLAS collaboration indicated that the threshold mass must be larger than 5.3 TeV \cite{atlas}.	
 	Hereupon, so far we have gained the ability to survey experimentally lengths with a scale up to two orders of magnitude smaller than the electroweak length scale, and no new physics, compared to the standard model, has been discovered with certainty.  	
  In Refs. \cite{das3,ali}, it has been shown that the GUP brings important qualitative and quantitative changes to
 		black hole thermodynamics. This causes the minimum energy for formation to be increased so that it could be pushed beyond the reach of LHC even if $M_S\sim 1$TeV.
 	Therefore, the length scale of new physics and gravity modifications shall be smaller than this scale. This implies that the best upper bound on $\beta_0$ which one can obtain potentially from LHC data should be of the order of $10^{30}$. 
 	
 	 Otherwise, considering the universality of the GUP, one can obtain an upper bound for $\beta_0$ from high-resolution measurements such as spectroscopy on the hydrogen atom, considering the ground state Lamb shift \cite{lamb}, the 1S-2S
 		level difference \cite{1s2s} and Landau levels using a scanning tunnel
 		microscope (STM) \cite{das,ett}. The values obtained by these methods in real conditions are greater than $10^{32}$. In the case of the upper bound obtained from the AURIGA detector \cite{AURIGA1,AURIGA2}, the situation is the same. However, the upper bound obtained from the lack of violation of the equivalence principle is in the order of $10^{21}$ \cite{ep}.

 Neutrino oscillation is a fantastic quantum phenomenon occurring at macroscopic distances. In fact, this process can be used to study quantum aspects in large distances even up to several hundred kilometers away \cite{blasone1,blasone2,ette2020,alok,Banerjee,naikoo,ette2022,ette2022D,li,jha,yadav,blasone3,ettef5}.
 The idea of neutrino oscillation was first put forward more than half a century ago and since then, using various neutrino sources, many experimental indications showing the transition between different neutrino flavors have been discovered \cite{historical}. Meanwhile, there are some debates on the basic issues of the theory of neutrino oscillations \cite{paradoxes}. At first glance and not so accurately, treating neutrinos as a plane wave or stationary state and assuming some additional and non-physical constraint such as equal either energy or momentum for mass eigenstates, one can obtain the probability of oscillation. However, a wave function with definite momentum spread throughout the space and a stationary  state does not depend on time. While neutrino oscillation is the result of dependence on spatial position and time evolution.  
 In order to provide a more consistent theoretical explanation for neutrino oscillation, we must take into account the localization of microscopic processes by which a neutrino is produced and detected based on the uncertainty principle and treat neutrinos as wave packets \cite{paradoxes,Nussinov,Kayser,giunti1,giunti2,Kiers,giunti3}. 
 
 An upper bound on the scale of a minimal length induced by a quantum gravity theory has already been obtained through neutrino oscillation study based on the stationary description which is inconsistent \cite{ Sprenger}. Meanwhile, in this paper, we will restudy this problem with taking into account the GUP to obtain the momentum wave functions of produced and detected neutrinos. The produced  state evolves with time in position space to reach the detector. Transformation of the momentum wave functions to the position wave functions is problematic; since there exists an measurable minimal length, $\Delta x_0=\sqrt{\beta}$, one cannot define a position eigenstate with a zero position uncertainty. Therefore, one needs to formulate appropriate Hilbert space representation of the modified Heisenberg algebra in quantum mechanics, which we call the GUP position eigenstate \cite{kempf}. Given that neutrino oscillation experiments are performed in energies very less than the Planck scale, the GUP induced effects are very small in comparison to the usual one so that we will keep the linear terms of $\beta$ and we obtain perturbatively the transition amplitude and consequently the transition probability.

 In principle, it is expected that the effects of the measurable minimal length occur in the Planck scale. However, the GUP is universal and similar to \cite{das,lamb,1s2s}, one can infer an upper bound on $\beta_0$ by neutrino oscillation experiments. For instance, we will use the  set up requirements of the far MINOS detector\cite{minos} to estimate an lower bound on $\beta_0$. 
 
 The paper is organized as follows: in the following section, we give a brief review of GUP. The oscillation probability is obtained in 
 section \ref{3}. We give a numerical analyzing in section \ref{4}. Finally, we summarize our results in section \ref{5}.

\section{GUP}
The Heisenberg algebra corresponding to the GUP given in Eq. (\ref{gup}) can be written as follows\footnote{Since detected neutrinos propagate along a macroscopic distance, we consider only one spatial dimension
	along the neutrino path.}: 
\begin{equation}
[{\bf x},{\bf p}]=i(1+\beta {\bf p}^2).\label{com.xp}
\end{equation}
In this paper, we consider the existence of a minimal position length and ignoring any restriction on the momentum space. Hence, in the momentum space, the operators ${\bf p}$ and ${\bf x}$ have the following representations:
\begin{equation}
{\bf p}\phi(p)=p\phi(p),\label{p}
\end{equation}
and 
\begin{equation}
{\bf x}\phi(p)=i(1+\beta p^2)\partial_p\phi(p).\label{x}
\end{equation}
This representation satisfies Eq. (\ref{com.xp}). A nonzero minimal uncertainty in the position implies that we cannot  have a physical state which is a position eigenstate \cite{kempf}. Nevertheless, 
let us use Eq. (\ref{x}) to obtain the position eigenstate in the momentum space
\begin{equation}
	i(1+\beta p^2)\partial_p\phi_x(p)=x\phi_x(p),\label{xx}
\end{equation}
where $x$ is the position eigenvalue. Because $\beta p^2\ll 1$ and we are interested in the first-order corrections, we can solve Eq. (\ref{xx}) by the perturbation method. Therefore, up to the first order of $\beta$,  the position eigenstate in the momentum space is given by:
\begin{equation}
	\phi_x(p)=\frac{1}{\sqrt{2\pi}}(1+\frac{i\beta}{3}xp^3)e^{-ipx}.\label{FP}
\end{equation}

%where $N$ is the normalization factor.
Furthermore, in order to preserve the symmetry property of ${\bf x}$ and ${\bf p}$, the identity operator in momentum space must be written as \cite{kempf}
\begin{equation}
{\bf 1}=\int \frac{dp}{1+\beta p^2}|p\rangle\langle p|.\label{id}
\end{equation}

 From Eq. (\ref{com.xp}), one can show that the momentum operator in position space is given by
\begin{equation}
{\bf p}\psi(x)=-i(1-\frac{\beta}{3}\partial_x^2)\partial_x\psi(x).
\end{equation} 
Accordingly, the modified Dirac equation in one dimension can be written as \cite{das2,pedram}
\begin{equation}
i\partial_0\psi=-\gamma_0\Big(i(1+\frac{\beta}{3}\partial_x^2)\gamma^1\partial_x+m\Big)\psi
\end{equation}
where $\gamma_0$ and $\gamma_1$ are two of Diac matrices. Therefore, the dispersion relation that results from this relation up to the first order of $\beta$ is written as follows:
\begin{equation}
E^2=p^2(1-\frac{2}{3}\beta p^2)+m^2.\label{DR}
\end{equation}
 
In order to explain neutrino oscillation in quantum mechanics framework, we must take into account the localization of microscopic processes by which a neutrino is produced and detected. The localization in quantum mechanics is based on the uncertainty principle. For each state in the representation of the Heisenberg algebra, the uncertainty principle can be deduced from the positivity of the norm 
\begin{equation}
||\Big({\bf x}-\langle {\bf x}\rangle +\frac{\langle [{\bf x},{\bf p}]\rangle}{2(\Delta p)^2}({\bf p}-\langle {\bf p}\rangle)\Big)||\geq 0.
\end{equation}
In fact, given that $\langle [{\bf x},{\bf p}]\rangle$ is imaginary, this relation is written as 
\begin{equation}
\langle\psi|({\bf x}-\langle {\bf x}\rangle)^2 +\Big(\frac{\langle [{\bf x},{\bf p}]\rangle}{2(\Delta p)^2}\Big)^2({\bf p}-\langle {\bf p}\rangle)^2|\psi\rangle\geq 0,
\end{equation}
which immediately implies that
\begin{equation}
\Delta x\Delta p\geq \frac{|\langle[{\bf x},{\bf p}]\rangle|}{2}.
\end{equation}
In the case of boundary of the physically allowed region, a state $|\psi\rangle$ must fulfill the following condition:
\begin{equation}
	\Big({\bf x}-\langle {\bf x}\rangle +\frac{\langle [{\bf x},{\bf p}]\rangle}{2(\Delta p)^2}({\bf p}-\langle {\bf p}\rangle)\Big)|\psi\rangle= 0,
\end{equation}
Using Eq. (\ref{com.xp}) and Eq. (\ref{x}), one casts the recent equation in the momentum representation as follows
\begin{equation}
\Big(i(1+\beta p^2)\partial_p-\langle {\bf X}\rangle+i\frac{1+\beta(\Delta p)^2+\beta \langle {\bf p}\rangle^2}{2(\Delta p)^2}(p-\langle {\bf p}\rangle)\Big)\phi(p)=0.\label{main}
\end{equation}

\section{Neutrino oscilation with GUP}\label{3}

 There are two approaches for explanation of neutrino oscillation; field theory and quantum mechanics approaches \cite{beuthe}.
In the field theory approach, neutrinos are treated as intermediate particles between the source and the detector, and the neutrino oscillation is the result of considering the details of the localization of particles participating in the production and detection processes. This leads to the uncertainty in determining the momentum of the mass eigenstates, such that a coherent superposition of the various mass eigenstate propagators contributes in the transition amplitude. Therefore, the oscillation pattern comes form the interference terms in the transition probability. In this approach, we should note that neutrinos are described by a Feynman propagator which is modified due the induced GUP corrections \cite{feynmanp}. Moreover, the interactions during the production and detection processes may be affected by the GUP due to the production of (microscopic) black hole productions which are predicted by ADD theory\cite{ed1}. A detailed study of this theory is beyond the scope of the present work and is left for future studies. 

In the case of quantum mechanical approach which is followed in the this paper, neutrinos are treated as  localized particles traveling between source and detector. Since the GUP affects a host of quantum mechanical problems, we must modify our considerations about the localization properties in this approach of the neutrino oscillation explanation as well. But the question may arise, why do we expect the uncertainty principle for neutrinos to be reformulated under quantum gravity as GUP? In the absence of a consistent theory to merge quantum mechanics with general relativity, we can use a semi-classical approximation through which the canonical commutation relations between the momentum operator $p^\mu$ and position operator $x^\mu$, which are $[x^\mu,p^\nu]=i\hbar\eta^{\mu\nu}$ in Minkowski space-time, in a curved
space-time with metric $g_{\mu\nu}$ have been generalized as
\begin{equation}\label{cc}
	[x^\mu,p^\nu]=i\hbar g^{\mu\nu}.
\end{equation}    
In Ref. \cite{metric}, by representing the position and momentum operators as covariant derivatives with an appropriate connection in the eight dimensional manifold, quantization has been geometrically interpreted as the curvature of phase space. Through this approach, implying the existence of an upper limit on the acceleration of particles moving along their worldlines, one can show that the corresponding metric tensor can be written as follows:
\begin{equation}
	g^{\mu\nu}\sim \left(1-c^4\frac{\ddot{x}^\sigma\ddot{x}_\sigma}{\cal A}\right)\eta^{\mu\nu},\label{gg}
\end{equation}
where $\ddot{x}^\sigma$ and $\cal A$ denote the four-acceleration which can be given in terms of momentum dispersion and the maximal acceleration, respectively. Substituting $g^{\mu\nu}$ from Eq. (\ref{gg}) in Eq. (\ref{cc})  and using the uncertainty relation $\Delta A\Delta B\le 1/2|\langle[A,B]\rangle|$, one can obtain GUP 
\begin{equation}
	\Delta x\Delta p\le\frac{\hbar}{2}+\frac{\alpha}{c^3}G\Delta p^2,
\end{equation}
where $\alpha$ is a free parameter. Moreover, as another viewpoint, the GUP can arise provided that relaxing the postulates of quantum mechanics, we ignore the position operator that satisfies the usual Heisenberg commutation relation \cite{kazemi}. Nevertheless, we base it on the universality of quantum gravity and exert the modifications due to GUP on the quantum mechanics approach of neutrino oscillation explanation.

The existence of minimal length has been explored previously using neutrino oscillation phenomenon with treating neutrinos as stationary\cite{Sprenger}. However, strictly speaking, the probability of finding a particle described by stationary states does not depend on time. 
 In quantum theory, propagating particles must be described generally by moving wave packets which fulfill the uncertainty principle deduced from Heisenberg algebra. As we saw, the quantum gravity induced minimal length leads to a modification of Heisenberg's algebra (Eq. \ref{com.xp}) and the corresponding uncertainty principle called GUP. Here, we will assume the neutrino wave packets satisfy the boundary allowed region of the GUP (Eq. \ref{main}).
 Accordingly, the scenario that we follow to obtain the probability of neutrino oscillation is as follows: using Eq. \ref{main} we must obtain the neutrino wave function in momentum representation at the source and detector. Then after temporal evolution of the source wave function, we must project the source and detector wave functions into the position representation. The amplitude of oscillation is equal to the spatial correlation of these wave functions. The squire of absolute value of the amplitude is the oscillation probability. Of course, we must keep in mind that the produced and detected neutrinos do not have a specific mass and they are said to be in the flavor eigenstate. Therefore, their states must be written as a superposition of all mass eigenstates.
So we begin by calculating the corresponding wave functions.

\subsection{Propagating neutrino wave function}
Our basis for calculating the related wave functions is Eq. (\ref{main}). In this equation, $\langle {\bf x}\rangle$ and $\langle {\bf p}\rangle$ are the mean value of the position and momentum operators, respectively. We chose  $\langle {\bf x}\rangle=0$ and $\langle {\bf p}\rangle=p_k$ in which $p_k$ is the average amount of momentum for the $k$'th mass eigenstate of neutrinos. Moreover, $\Delta p$ is taken the momentum uncertainty of neutrino during the creation process and we denote it by ${\sigma_{p}}_P$. Since we are interested in the GUP corrections up to the first order of $\beta$, we solve Eq. (\ref{main}) by the perturbation method. In this manner, the created momentum wave function of neutrinos in $k$'th  mass eigenstate is obtained as follows:
\begin{equation}
	\phi^\text{P}_k(p;p_k,{\sigma_p}_\text{P})=\frac{e^{-\frac{(p-p_k)^2}{4 {\sigma_p^2}_\text{P}}}}{ \sqrt[4]{2 \pi \sigma_{pP}^2} } \left(1-\frac{1}{4} \beta  (p-p_k)^2+\frac{\beta}{24{\sigma_p^2}_\text{P}}  (p-p_k)^3 (3 p+5p_k)-\frac{1}{8} \beta  {\sigma_p^2}_\text{P}\right),\label{phip}
\end{equation} 
 which is normalized up to the first order of $\beta$. 
 What participates in weak interactions are the flavor eigenstates of neutrinos, which are written in terms of mass eigenstates as follows:
 \begin{equation}
 	|\nu_\alpha\rangle=\sum_{k=1}^3U^*_{\alpha k}|\nu_k\rangle,
 \end{equation}
 where $U_{\alpha k}$ denotes the mixing unitary matrix elements.
 Therefore, using Eq. (\ref{FP}), Eq. (\ref{id}) and Eq. (\ref{DR}), one can give the time evaluated of the neutrino wave function with definite flavor in the Schrodinger picture by
 \begin{equation}
\psi^{\text{P}}_\alpha(x,t)=\frac{1}{\sqrt{2\pi}}\sum _{k=1}^3 U^*_{\alpha k}\int \frac{dp}{1+\beta p^2} \phi^\text{P}_k(p;p_k,{\sigma_p}_\text{P})(1+\frac{i}{3}\beta xp^3)\text{e}^{-i\left(E_k(p)t-px\right)},\label{psip}
 \end{equation}
 where $E_k(p)=\sqrt{p^2(1-\frac{2}{3}\beta p^2)+m_k^2} \approx E_k^0(p)-\frac{1}{3}\frac{\beta p^4}{E_k^0(p)}$ in which $E_k^0(p)=\sqrt{p^2+m_k^2}$. We assume the momentum wave functions, Eq. (\ref{phip}), are sharply peaked around the corresponding average momentum, i.e. ${\sigma_p}_\text{P}\ll p_k$. Under this condition, we can keep the terms which are liner with respect to $\beta$ for energies below the minimal length scale. Moreover,  $E_k^0(p)$ can be approximated by $E_k^0(p)\approx E_k-v_k(p-p_k)$ where $E_k=\sqrt{p_k^2+m_k^2}$ and $v_k=p_k/ E_k$ is the  group velocity of each wave packet with definite mass. The other point which should be noted is that due to the large energy scale for the emergence of the effects of the minimal measurable length and due to the limited neutrino energy at different sources, we can use the usual uncertainty relation, ${\sigma_x}_P{\sigma_p}_P=1/2$, to replace momentum uncertainty with position uncertainty. After taking into account all these issues, the wave function of the propagated neutrinos up to the first order of $\beta$ is obtained as follows:
 %\begin{eqnarray}
%\psi^{\text{P}}_\alpha(x,t)=\frac{1}{{96 \sqrt[4]{2 \pi {\sigma_x}_P^2} {\sigma_x}_P^6}}\sum _{k=1}^3 U^*_{\alpha k} e^{-i (E_k t- p_k x)-\frac{(x-t v_k)^2}{4 {\sigma_x}_P^2}} \{32 {\sigma_x}_P^6 \left(3-\beta  p_k^2 (3-i p_k (x-t v_k))\right)\nonumber\\
%-3 \beta  {\sigma_x}_P^4 (9-16 p_k (x-t v_k) (i-p_k (x-t v_k)))+10 i \beta  {\sigma_x}_P^2 (x-t v_k)^2 (3 i-4 p_k (x-t v_k))+7 \beta  (x-t v_k)^4\}.\label{psixp}
% \end{eqnarray}

\begin{eqnarray}
	\psi^{\text{P}}_\alpha(x,t)=\frac{1}{ \sqrt[4]{2 \pi \sigma_{xP}^2}} \sum _{k=1}^3 U^*_{\alpha k}\text{e}^{-i( E_k t+p_k x)-\frac{(x-v_kt)^2}{4 \sigma_{xP}^2}} \bigg[1+\beta  \bigg(&&\hspace{-0.4cm}\frac{i(x-v_kt)}{\sigma_{xP}^2}-\frac{5(x-v_k)^2}{16\sigma_{xP}}-\frac{ip_k (x-v_kt)^3}{6 \sigma_{xP}^4}\nonumber\\
	&&+\frac{(x-v_kt)^4}{32 \sigma_{xP}^6}+\frac{7}{32 \sigma_{xP}^2}\bigg)\bigg].\label{psixp}
\end{eqnarray}

\subsection{Detected neutrino wave function}
The wave function of the detected neutrino must be localized around the detector position, a distance $L$ from the source. In addition, we assume that the mean momentum of the detected neutrino is equal to the mean momentum of the propagated neutrino due to the momentum conservation. Therefore, in Eq. (\ref{main}) for detected neutrino, we take $\langle {\bf x}\rangle=L$ and $\langle {\bf p}\rangle=p_k$. Similar to the previous case, the wave function of detected neutrinos which is the solution of Eq. (\ref{main}) is given by
\begin{eqnarray}
	\phi^\text{D}_k(p;p_k,{\sigma_p}_\text{D})=\frac{1}{{\sqrt[4]{2 \pi {\sigma_p}_\text{D}^2}}}\text{e}^{-\frac{(p-p_k)^2}{4 {\sigma_p}_\text{D}^2}-i p L} \left[1+\frac{\beta}{24}\left(-3   {\sigma_p}_\text{D}^2+2 \left(-3 (p-p_k)^2+4 i p^3 L\right)+\frac{(p-p_k)^3 (3 p+5 p_k)}{{\sigma_p}_\text{D}^2}\right)\right].\nonumber\\
\end{eqnarray}
This wave function does not represent a propagating neutrino, therefore, it does not evolve in time.
Consequently, using Eq. (\ref{FP}) and Eq. (\ref{id}), one can write the position wave function of the detected neutrinos as follows:
\begin{equation}
	\psi^{\text{D}}_\alpha(x)=\frac{1}{\sqrt{2\pi}}\sum _{k=1}^3 U^*_{\alpha k}\int \frac{dp}{1+\beta p^2} \phi^\text{D}_k(p;p_k,{\sigma_p}_\text{D})(1+\frac{i}{3}\beta xp^3)\text{e}^{ipx}.\label{psid}
\end{equation}
With assumption similar to what considered for Eq. (\ref{psixp}), up to the first order of $\beta$, one can simplify the position wave function of the detected neutrinos as follows:
\begin{eqnarray}
\psi^{\text{D}}_\alpha(x,t)=\frac{1}{24\sqrt[4]{2 \pi {\sigma_x}_\text{D}^2}}\sum _{k=1}^3 U^*_{\alpha k}e^{i(x-L)p_k- \frac{(x-L)^2}{4 {\sigma_x}_\text{D}^2}}\Big(&&\hspace{-0.4cm}24+8 i \beta  L p_k^3+\frac{12 \beta  L p_k^2 (L-x)}{{\sigma_x}_\text{D}^2}\nonumber\\
&&\hspace{-0.7cm}-\frac{2 i \beta  p_k \left(L^3+6 L {\sigma_x}_\text{D}^2-3 L x^2+2 x^3-12 {\sigma_x}_\text{D}^2 x\right)}{{\sigma_x}_\text{D}^4}\nonumber\\
&&\hspace{-0.7cm}-\frac{\beta  \left(6 {\sigma_x}_\text{D}^2 \left(L^2-6 L x+5 x^2\right)+(L-x)^3 (L+3 x)-21 {\sigma_x}_\text{D}^4\right)}{4 {\sigma_x}_\text{D}^6}\Big).\label{psixd}
\end{eqnarray}

\subsection{Oscillation probability}
The oscillation amplitude  is obtained through calculating the correlation of the wave functions given by Eqs. (\ref{psixp}) and (\ref{psixd})
\begin{equation}
	A_{\alpha\beta}=\int{\psi^D_\beta}^*(x,t)\psi^P_{\alpha}(x,t)dx,
\end{equation}

and consequently the transition probability is given by
\begin{equation}
	P_{\alpha\beta}=|A_{\alpha\beta}|^2.
\end{equation}
Because the neutrino emission and arrival time are not measured, we should integrate over the time during which a neutrino travels from the source to the detector. Hence, by direct calculation, one can obtain the transition probability up to the first order of $\beta$ as follows:

\begin{equation}
P_{\alpha\beta}\simeq\sum_i\sum_j  U_{\alpha i}U^*_{\beta i} U^*_{\alpha j}U_{\beta j}\text{e}^{-2\pi i\frac{L}{L^{\text{osc}}_{ij}}-\frac{L^2}{{L^{\text{coh}}_{ij}}^2}-2\pi^2\rho^2\frac{\sigma_x^2}{{L^{\text{osc}}_{ij}}^2}}f_{ij}(\beta),\label{palphabeta}
\end{equation}
where
\begin{equation}
	f_{ij}(\beta)=1-\beta\left(\frac{i \pi \left(4 E^2\sigma_x^2+1\right)}{2 \sigma_x^2}\frac{L}{L^{\text{osc}}_{ij}}+\frac{ \left(8E^2\sigma_x^2-5\right)}{4 \sigma_x^2}\frac{L^2}{{L^{\text{coh}}_{ij}}^2}+\frac{i \pi ^3   }{6E^2\sigma_x^4}\frac{L^3}{{L^{\text{osc}}_{ij}}^3}+\frac{3}{4 \sigma_x^2}\frac{L^4}{{L^{\text{coh}}_{ij}}^4}\right),\label{fij}
\end{equation}
with $L^{\text{osc}}_{ij}=(4\pi E)/\Delta m_{ij}^2$ and $L^{\text{coh}}_{ij}=(4\sqrt{2}\sigma_xE^2)/|\Delta m^2_{ij}|$.
$\sigma_x$ is an effective width of the shape factor of the transition amplitudes.  Therefore, it depends on the widths of both the production and detection wave packets ${\sigma_x}_P$ and ${\sigma_x}_D$. For the conditions assumed here, establishing the Eq. (\ref{main}) in both production and detection processes, $\sigma_x$ is given by the following relation \cite{paradoxes}:
\begin{equation}
	\sigma_x=\sqrt{{\sigma_x}_P^2+{\sigma_x}_D^2}.\label{sigmax}
\end{equation}
Hence, it is of the order of the largest between ${\sigma_x}_P$ and ${\sigma_x}_D$.

In this calculation, we use the following statements:
\begin{itemize}
	\item We assume that neutrinos are extremely relativistic. Therefore, their mass eigenstate energy and momentum are approximated by the following relations:
	\begin{equation}
	E_i=E+\rho\frac{m_i^2}{2E},
	\end{equation}
and
\begin{equation}
	p_i=E-(1-\rho)\frac{m_i^2}{2E},
\end{equation}
where $E$ is the neutrino energy in the limit of zero mass and $\rho$ is a dimensionless quantity that can be estimated from energy-momentum conservation in the production processes \cite{rho}. 	
	\item In the limit of extremely relativistic, the group velocity can be estimated as follows:
	\begin{equation}
	v_j\simeq 1-\frac{m_j^2}{2E}.	
	\end{equation}
\item Up to first order of $\beta$, the transition probability is dependent on the propagation length $L$ maximum up to the forth power of it. For every power of $L$, we keep those terms which are dependent on the neutrino mass upmost in the order to the twice of the $L$ power. 

\end{itemize}
%\begin{equation}
%P_{\alpha\beta}\propto \sum_j\sum_k  U_{\alpha j}U^*_{\beta j} U^*_{\alpha k}U_{\beta k}e^{-\frac{2i\pi L}{L^{osc}_{ij}}LF_1(\beta)-\frac{L^2}{{L^{coh}_{ij}}^2}F_2(\beta)}{\cal F}_{ij},
%\end{equation}
%where
%\begin{equation}
%F_1(\beta)\equiv 1-\frac{(-12+\xi)(m_i^2+m_j^2)^2}{32E^4\sigma_x^4}L^2\beta+\frac{-9+4E^2\sigma_x^2}{4\sigma_x^2}\beta+...\hspace{1mm},
%\end{equation}
%\begin{equation}
%F_2(\beta)\equiv 1-\frac{-7+4E^2\sigma_x^2}{2\sigma_x^2}\beta+\frac{L^2}{{L^{coh}_{ij}}^2}\frac{\beta}{\sigma_x^2}+...\hspace{1mm},
%\end{equation}
%\begin{equation}
%{\cal F}_{ij}\equiv\exp\Big[-2\pi^2(1-\xi)^2\big(\frac{\sigma_x}{L_{ij}^{osc}}\big)^2-2(m_i^2+m_j^2)(-1+\xi)\beta\Big]
%\end{equation}
\section{Numerical analysis}\label{4}
Although quantum gravity has been of interest to scientists for several decades, no significant success has been achieved in establishing a precise and definitive theory. The reason is that the effects of quantum gravity usually appear at the Planck scale, an energy scale beyond the reach of any kind of experiment. However, as it was said, in order to construct a consistent quantum gravity theory, it is necessary to add new fundamental principles such as the existence of minimal measurable length, space-time noncommutativity, etc. to our fundamental principles. As a rule, these new principles are universal and they can affect theories that can be investigated with the available experiments. For instance, the neutrino oscillation theory is affected by the GUP as it was studied in the previous section.
In this section, we want to give a numerical estimation for the GUP modifications to the neutrino oscillation probability obtained in the previous section.

As one can see from Eqs. (\ref{palphabeta}) and (\ref{fij}), the GUP modifications affects on both the oscillation wave length and the decoherence factors. We take the following values for the usual neutrino oscillation parameters \cite{oscillation parameters}:
\begin{eqnarray}
	 \Delta m_{21}^2\approx7.39\times10^{-5}eV^2;\hspace{1cm}\Delta m_{31}^2\approx2.45\times10^{-3}eV^2;\nonumber\\
	\theta_{12}\approx33.82;\hspace{1cm}\theta_{13}\approx8.61;\hspace{1cm}\theta_{23}\approx49.7.
\end{eqnarray}
The other effective parameter that plays an important role in the magnitude of the GUP modifications is $\sigma_x$. As it can be seen from Eqs. (\ref{palphabeta}) and (\ref{fij}), the smaller $\sigma_x$ leads to the more visible effects of the GUP modifications. However, it should be noted that in order to see the phenomenon of neutrino oscillation, $\sigma_x$ should not be so small that the coherence length becomes smaller than the oscillation length. For a fixed teravel distance, the GUP modifications increase with the amount of neutrino energy. If the energy of neutrinos is in order of $1\text{GeV}$ like MINOS experiment in which both production and detection processes are performed through nuclear reactions, it is natural to take $\sigma_x\sim 10^{-15}\text{m}$. Comparing with the MINOS results, we estimate an upper bound on the GUP parameter $\beta$. In this experiment, the energy of the neutrinos varies from several hundreds MeV to several tens GeV. In this energy interval, the coherence length is certainly larger than the oscillation one.

\begin{figure}
	\centering
	\includegraphics[scale=0.9]{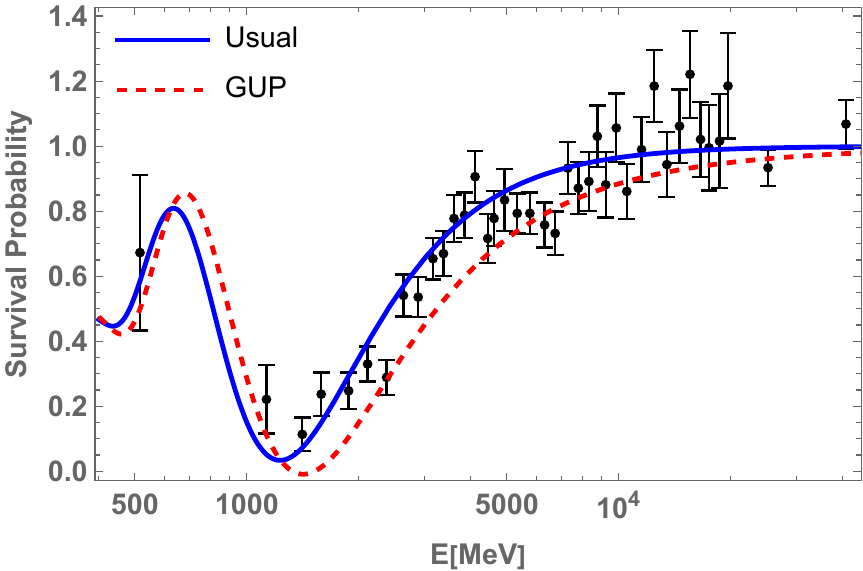}
	\caption{Survival probability in terms of neutrino energy for initial muon neutrino with fixed travel distance, $L=735\text{km}$. Blue (dashed) and red (dashed) lines illustrate the theoretical predictions based on usual and generalized uncertainty	principle (GUP), respectively. We also show the data of MINOS data (black, circle) taken from Ref. \cite{minos}.}\label{f1} 
\end{figure}

In the MINOS experiment, the distance between the neutrino source and the far detector is 735 km. Taking $\beta_0= 5\times 10^{25}$, we compare the long range survival probability for an initial muon neutrino in theories both with and without GUP modifications through red (dashed) and blue (solid) line, respectively, in Fig. \ref{f1}. The data of MINOS experiment (black, circle) is also shown in this figure. Therefore, the accelerator neutrino oscillation experiments such as MINOS with more precision can provide a more suitable opportunity to study the GUP.

%Therefore, the upper bound, which one can obtain from the MINOS experiment on the GUP deformation parameter $\beta_0$, will be much stronger than the corresponding one which can be obtained potentially from weak interactions (for example, from the LHC experiment). 
\section{Summary and conclusions}\label{5}
Although experimental research of quantum gravity is far from available, its effects are universal and can be investigated in existing experiments. The existence of the measurable minimal length is one of the effects of quantum gravity that can affect low energy physics and provide us with the possibility to study experimentally its effects \cite{das}. In fact, due to quantum gravity it is no longer possible to spatially localize a wave function to arbitrary precision and it leads to GUP given in Eq. (\ref{gup}). On the other hand, to give an appropriate and accurate explanation of the neutrino oscillation phenomena, one needs to consider the localization properties of neutrinos during production, propagation and detection processes \cite{paradoxes}. The wave packet approach in quantum mechanics provides us with a convenient framework for considering this issues \cite{Kayser}. In this paper, we have obtained the GUP effects on the neutrino oscillation via wave packet approach. 
Given that the GUP induce an measurable minimal length, $\Delta x_0=\sqrt{\beta}$, one cannot define a position eigenstate with a zero position uncertainty. Therefore, using the method developed in Ref. \cite{kempf}, we have formulated appropriate Hilbert space representation to construct the produced and detected neutrino state in the position representation. Meanwhile, since the GUP effects are expected to be very small at the energy scale at which neutrino oscillation experiments are performed, we have treated these effects perturbatively and obtained the corrections on the oscillation probabilities at the leading order. 

Next, we performed a numerical analysis on the GUP corrections, which shows that they depend on the effective width of the shape factor of the transition amplitudes, $\sigma_x$; smaller $\sigma_x$ makes these corrections more prominent. Therefore, through neutrino oscillation experiments with energies of about a few GeV such as the MINOS experiment in which one can take $\sigma_x\sim 10^{-15}\text{m}$, we can obtain a  stronger upper bound on the GUP deformation parameter $\beta_0$ in comparison to other experiments. In particular, through Fig. \ref{f1}, we have illustrated a deviation due to GUP modifications from the standard theoretical predictions for the survival probability of muon neutrinos by taking $\beta_0= 5\times 10^{25}$. The comparison with MINOS experimental data shows that if the length scale of the GUP is of the order of $l_{GUP}\sim10^{13}l_{Pl}$ where $l_{Pl}\equiv1.6\times10^{-35}\text{m}$, by measuring and analyzing the data from these types of experiments more precisely, the effects of the measurable minimal length can be explored.
 Hence, in this way, one can obtain a very stronger upper bound on $\beta_0$ compared to the bounds obtained in Refs. \cite{lamb,1s2s,das,ett,AURIGA1,AURIGA2}. But the bound obtained from the effect of GUP on the equivalence principle \cite{ep} is stronger than this value.

%\section*{Acknowledgement}
% The author would like to thank the Research Deputy of the University of Qom for support.


\begin{thebibliography}{99}
\bibitem{gravity1}
P.K. Townsend, 
 Phys. Rev. {\bf D 15}, 2795 (1976).

\bibitem{gravity2}
D. Amati, M. Ciafaloni, and G. Veneziano, 
Phys. Lett. {\bf B 216}, 41 (1989).

\bibitem{gravity3}
 L.J. Garay, 
  Int. J. Mod. Phys. {\bf A 10}, 145 (1995).

\bibitem{das}
S. Das and E.C. Vagenas, Phys. Rev. Lett. {\bf 101}, 221301 (2008).

\bibitem{GUP1}
D. Amati, M. Ciafaloni, G. Veneziano, Phys. Lett. {\bf B 216}
 41 (1989).

\bibitem{GUP2} 
 M. Maggiore, Phys. Lett. {\bf B 304},  65 (1993)
[arXiv:hep-th/9301067].

\bibitem{GUP3} 
L. J. Garay, Int. J. Mod. Phys.
{\bf A 10},  145 (1995) [arXiv:gr-qc/9403008].

\bibitem{ed1}
N. Arkani-Hamed, S. Dimopoulos and G. R. Dvali, Phys. Lett. {\bf B429}, 263 (1998).

\bibitem{ed2}
I. Antoniadis, N. Arkani-Hamed, S. Dimopoulos et al., Phys. Lett. {\bf B436}, 257-263 (1998). 

\bibitem{ed3}
L. Randall, R. Sundrum, Phys. Rev. Lett. {\bf 83}, 3370-3373 (1999).

\bibitem{ed4} 
L. Randall, R. Sundrum, Phys. Rev. Lett. {\bf 83}, 4690-4693 (1999).

\bibitem{bh1}
 S.B. Giddings, S.D. Thomas, Phys. Rev. D {\bf 65}, 056010 (2002).

\bibitem{bh2}
S. Dimopoulos, G.L. Landsberg, Phys. Rev. Lett. {\bf 87}, 161602 (2001).

\bibitem{bh3}
N. Arsene, R. Casadio, and O.  Micu, Eur. Phys. J. C {\bf 76}, 384 (2016).

\bibitem{cms}
S. Chatrchyan et al., JHEP {\bf 07}, 178 (2013).

\bibitem{atlas}
G. Aad et al., Phys. Rev. Lett. {\bf 112}, 091804 (2014).

\bibitem{das3}
M. Cavaglia, S. Das, R. Maartens, Classical and Quantum Gravity {\bf 20}, L205 (2003).

\bibitem{ali}
A. F. Ali, Journal of High Energy Physics {\bf 12}, 1-14 (2012).

\bibitem{lamb}
A. F. Ali, S. Das, E. C.  Vagenas, Phys. Rev. D {\bf 84}, 044013 (2011)

\bibitem{1s2s}
C. Quesne, and  V. M. Tkachuk, Phys. Rev. A {\bf 81}, 012106 (2010).

\bibitem{ett}
M.M. Ettefaghi, S.M. Fazeli, Phys. Rev. Lett. {\bf 104}, 119001 (2010).

\bibitem{AURIGA1}
 F. Marin, and et al., Nature Phys. {\bf 9}, 71 (2013).

\bibitem{AURIGA2}
F. Marin,  et al., New J. Phys. {\bf 16}, 085012 (2014).

\bibitem{ep}
S. Ghosh, Class. Quantum Grav. {\bf 31}, 025025 (2014).

\bibitem{blasone1}
 M. Blasone, F. Dell’Anno, S. De Siena, F. Illuminati, Europhys.
Lett. \textbf{85}, 50002 (2009).

\bibitem{blasone2}
M. Blasone, F. Dell’Anno, S. De Siena, F. Illuminati, Europhys.
Lett. \textbf{112}, 20007 (2015).

\bibitem{ette2020}
M.M. Ettefaghi, Z.S. Tabatabaei Lotfi, and R. Ramezani Arani,
Europhys. Lett. \textbf{132}, 31002 (2020).

\bibitem{alok}
 A.K. Alok, S. Banerjee, S.U. Sankar, Nucl. Phys. \textbf{B 909}, 65 (2016).
 
\bibitem{Banerjee}
S. Banerjee, A.K. Alok, R. Srikanth, B.C. Hiesmayr, Eur. Phys. J.
\textbf{C 75}, 1 (2015).

\bibitem{naikoo}
J. Naikoo, A.K. Alok, S. Subhashish Banerjee, U. Sankar, G.
Guarnieri, C. Schultze, B.C. Hiesmayr, Nucl. Phys. \textbf{B 951}, 114872
(2020).

\bibitem{ette2022}
Z. Askaripour Ravari, M. M. Ettefaghi, S. Miraboutalebi, Eur. Phys.
J. Plus \textbf{137}, 488 (2022).


\bibitem{ette2022D} M.M. Ettefaghi, R. Ramazani Arani, Z.S. Tabatabaei Lotfi, Phys. Rev. D \textbf{105}, 095024 (2022).

\bibitem{li}
Y.W. Li, L.J. Li, X.K. Song, D. Wang, L. Ye, Eur. Phys. J. \textbf{C 82}, 799 (2022).

\bibitem{jha}
A.K. Jha, A. Chatla, Eur. Phys. J. Spec. Top.\textbf{ 231}, 141 (2022).

\bibitem{yadav}
13. B. Yadav, T. Sarkar, K. Dixit, A.K. Alok, Eur. Phys. J. \textbf{C 82}, 1
(2022).

\bibitem{blasone3}
M. Blasone, S. De Siena, and C. Matrella,% Wave packet approach to quantum correlations in neutrino oscillations,
Eur. Phys. J. C, \textbf{81}, 660 (2021).

\bibitem{ettef5}
M.M. Ettefaghi, and Z. Askaripour Ravari,% "Quantum coherence and entanglement in neutral-current neutrino oscillation in matter." 
Eur. Phys. J. C \textbf{83}  417 (2023).

\bibitem{historical}
S. M. Bilenky, Phys. Scripta {\bf 121}, 17 (2005).

\bibitem{paradoxes}
E. Kh. Akhmedov, and A. Yu. Smirnov, Phys. Atom. Nucl. {\bf 72}, 1363 (2009).

%\bibitem{glashow}
%A.G. Cohen, S.L. Glashow,  and Z. Ligeti, %Disentangling neutrino oscillations. 
%Phys. Lett. B \textbf{678}, 191 (2009). 

\bibitem{Nussinov}
S. Nussinov,% "Solar neutrinos and neutrino mixing." 
Phys. Lett. B \textbf{63}, 201 (1976).

\bibitem{Kayser}
 B. Kayser, Phys. Rev. {\bf D 24}, 110 (1981).
 
 \bibitem{giunti1}
C. Giunti, C.W. Kim, and U.W. Lee, Phys. Rev. {\bf D 44}, 3635 (1991).

\bibitem{giunti2}
C. Giunti, C. W. Kim and U. W. Lee, Phys. Lett. {\bf B 274}, 87 (1992).

\bibitem{Kiers}
K. Kiers, S. Nussinov, and N. Weiss, %. Coherence effects in neutrino oscillations. 
Phys. Rev. D \textbf{53}, 537 (1996).

\bibitem{giunti3}
C. Giunti, and C.W. Kim, % Coherence of neutrino oscillations in the wave packet approach. 
Phys, Rev, D \textbf{58}, 017301 (1998).

\bibitem{Sprenger}
M. Sprenger, P. Nicolini, and M. Bleicher, %Neutrino oscillations as a novel probe for a minimal length. 
Classical and Quantum Gravity, \textbf{28}, 235019 (2011).

\bibitem{kempf}
A. Kempf, G. Mangano, and R. Mann, Phys. Rev. D52, 1108 (1995).

\bibitem{minos}
A. B. Sousa (MINOS and MINOS+ Collaborations), AIP Conf.
Proc. \textbf{1666}, 110004 (2015).

\bibitem{das2}
 S. Das, E.C. Vagenas, and A.F. Ali, Phys. Lett. {\bf B 690},  407 (2010).
 
 \bibitem{pedram}
P. Pedram,  Phys. Lett. {\bf B 702}, 295 (2011).


\bibitem{beuthe}
M. Beuthe, Physics Reports, {\bf 375}, 105 (2003).

\bibitem{feynmanp}
R. Casadio, F. Wenbin, K. Ibere, and S. Fabio, Phys. Lett. {\bf B 838}, 137722 (2023).

\bibitem{metric}
 S. Capozziello, G. Lambiase, and G. Scarpetta, International Journal of Theoretical Physics \textbf{39}, 15 (2000).
 
\bibitem{kazemi}
M.J. Kazemi, and G. Jafari, arXiv preprint arXiv:2402.11350 (2024).

%\bibitem{coherence}
%E. Kh. Akhmedov, D. Hernandez, and A. Yu. Sminrnov, JHEP {\bf 1204}, 052 (2012).

\bibitem{rho}
C. Giunti, J. High Energy Phys. {\bf 0211}, 017 (2002).

\bibitem{oscillation parameters}
I. Esteban, M.C. Gonzalez-Garcia, A. Hernandez-Cabezudo, M. Maltoni, T. Schwetz, JHEP {\bf 01}, 106 (2019).
\end{thebibliography}
\end{document}